\documentclass[letterpaper, 10 pt, conference]{ieeeconf}  

\IEEEoverridecommandlockouts                              

\usepackage{graphicx} 
\usepackage{amssymb}
\usepackage{amsmath}

\title{\LARGE \bf
Nonlinear Model Predictive Control for Leaderless UAV Formation Flying with Collision Avoidance under Directed Graphs}

\author{	Yiming Wang,
Yao Fang,
Jie Mei, \emph{Member},
Youmin Gong, \emph{Member},
Guangfu Ma
\thanks{This work was supported in part by Guangdong Basic and Applied
Basic Research Foundation under Grant 2023B1515120018 and Grant
2024B1515040008, in part by the Shenzhen Fundamental Research Program under Grant JCYJ20241202124010014, and in part by the National Natural Science Foundation
of China under Grant 62073098. (Corresponding author: Jie Mei.)}
\thanks{The authors are with the School of Intelligence
Science and Engineering and Guangdong Key Laboratory
of Intelligent Morphing Mechanisms and Adaptive Robotics, Harbin
Institute of Technology (Shenzhen), Shenzhen 518055, China
(e-mail: 22b953010@stu.hit.edu.cn;
 23s053067@stu.hit.edu.cn; jmei@hit.edu.cn; gongyoumin@hit.edu.cn; magf@hit.edu.cn)}}%

\begin{document}

\maketitle
\thispagestyle{empty}
\pagestyle{empty}

\begin{abstract}

This paper studies the leaderless formation flying problem with collision avoidance for a group of unmanned aerial vehicles (UAVs), which requires the UAVs to navigate through cluttered environments without colliding while maintaining the formation. The communication network among the UAVs is structured as a directed graph that includes a directed spanning tree. A novel distributed nonlinear model predictive control (NMPC) method based on the model reference adaptive consensus (MRACon) framework is proposed. Within this framework, each UAV tracks an assigned reference output generated by a linear reference model that utilizes relative measurements as input. Subsequently, the NMPC method penalizes the tracking error between the output of the reference model and that of the actual model while also establishing constraint sets for collision avoidance and physical limitations to achieve distributed and safe formation control. Finally, simulations and hardware experiments are conducted to verify the effectiveness of the proposed method. 
\end{abstract}

\section{INTRODUCTION}

In recent years, with the rapid development of unmanned systems technology, unmanned aerial vehicles (UAVs) 
have been deployed in various tasks, such as cooperative exploration \cite{zhou2021fuel}, environmental mapping \cite{xu2022omni}, search and rescue \cite{wu2023adaptive}, collaborative payload transportation \cite{wehbeh2020distributed},
  and so on. Compared to the individual UAV working alone, multiple 
 UAVs can accomplish complex tasks more efficiently, thus receiving wide attention from scholars \cite{javaid2023communication,hegde2021multi}. 
 Formation flying control is essential for the coordinated control of UAV swarms. 
 Considering the presence of obstacles in the flying
  environment, the problem of formation flying with collision avoidance has been proposed.
  This problem involves two key issues: first, UAVs take off from arbitrary initial positions, form a formation, and maintain
   this formation while maneuvering at a set speed; second, the UAVs must avoid collisions with obstacles and other UAVs in the formation.

 Currently, multi-UAV formation control methods primarily include leader-following control \cite{zhang2020multi}, consensus-based control \cite{wu2020new}, artificial potential function control \cite{pan2021improved}, and angle constraint control \cite{chen2023angle}. Among them, consensus-based control has the advantages of distributed execution, lower communication volume, and robustness, making it suitable for large-scale multi-agent systems.
However, due to safety requirements for avoiding collisions, limitations in the flying area and velocity, as well as control input constraints, ordinary consensus-based control cannot effectively achieve the UAV formation flying mission with collision avoidance.

  Model predictive control (MPC) can explicitly handle various constraints and perform optimal control,
   making it suitable for addressing the UAV flight control problem with collision avoidance constraints \cite{liu2023integrated}.
   Moreover, MPC has been extended to multi-robot systems, wherein researchers have implemented distributed model predictive control (DMPC) to accomplish a point-to-point transfer task \cite{luis2019trajectory}.
   In these studies, the control problem is formulated as a quadratic programming problem, where the dynamics model of UAVs is simplified to a linear one, and linear inequality constraints are set to avoid collisions.
	However, the linearization of the dynamic model leads to a degradation in both control accuracy and robustness for the UAV.
	The nonlinear model predictive control (NMPC) method has garnered increasing attention in unmanned aerial vehicles (UAVs) \cite{lindqvist2020nonlinear}. 
	There are two primary reasons for this interest: firstly, NMPC can generate collision-free trajectories and corresponding control 
	input sequences purely using the UAV's nonlinear dynamics model; secondly, it can handle non-convex constraints, such as collision avoidance constraints, more directly than linear MPC.

                

	A model predictive contouring control (MPCC) method is proposed for unmanned surface vehicle formation \cite{zhao2024distributed}, transforming the formation problem into a cooperative path-tracking problem.
	This method depends on the generation of a centralized reference trajectory.
	A fast-distributed MPC method is proposed for UAV formation \cite{liu2018distributed}, which brings the UAVs to a predefined submanifold.
	In the method, the communication topology of UAVs is designed as an undirected cycle graph.
	Furthermore, implementing UAV formation with collision avoidance under a directed graph remains challenging.
	Recently, a unified framework called model reference adaptive consensus (MRACon) has been proposed \cite{mei2021unified},
 which provides a simple yet efficient solution to the problem of achieving consensus in uncertain multi-agent systems under directed graphs.
 This framework is similar to the model reference adaptive control (MRAC) concept, where a tracking controller is designed to follow the output of the reference model.
 In the MRACon framework, a consensus algorithm using relative measurements is designed to generate the reference model output, and each agent is assigned a reference output to track.            
  Inspired by the MRACon framework, we propose a novel distributed NMPC method to address the problem of UAV formation flying with collision avoidance under directed graphs.

The contributions are summarized as:
\begin{itemize}
\item A distributed formation control algorithm employing NMPC to prevent collisions among neighboring UAVs and obstacles.
\item An efficient approach for constructing reciprocal collision avoidance constraints by predicting the trajectories of neighboring UAVs.
\item Simulations and real-world experiments are conducted to validate its effectiveness.
\end{itemize}


  \section{PRELIMINARIES AND PROBLEM FORMULATION}
  
  \subsection{UAV Dynamics}
  We consider that  $n$ ($n>$2) UAVs are in the formation, labeled as $i$=1,...,$n$.
 For clarity, we denote the $i$th UAV using a superscript $i$.
   A cascaded control structure is employed for UAV flight, where a first-order dynamic model approximates the motions of the roll and pitch angles \cite{lindqvist2021reactive}.
   The state of the $i$th UAV is denoted as $\boldsymbol{x}^i=[\boldsymbol{p}^i,\boldsymbol{v}^i, \phi^i,\theta^i, \psi^i]^{\mathrm{T}}\in\mathbb{R}^9$, where $\boldsymbol{p}^i=[p_x^i, p_y^i, p_z^i]^{\mathrm{T}}\in\mathbb{R}^3$, $\boldsymbol{v}^i=[v_x^i, v_y^i, v_z^i]^{\mathrm{T}}\in\mathbb{R}^3$ represent the position and velocity of the $i$th UAV.
   $\phi^i, \theta^i, \psi^i \in \mathbb{R}$ are the roll, pitch, and yaw angles.
The control input is $\boldsymbol{u}^i=[T^i, \phi_{ref}^i, \theta_{ref}^i, \psi_{rate}^i]^{\mathrm{T}}\in\mathbb{R}^4$, in which $\phi_{ref}^i, \theta_{ref}^i \in \mathbb{R}$ are the command of the roll and pitch angles, $\psi_{ref}^i \in \mathbb{R}$ is the command rates of the yaw angle, $T^i \in \mathbb{R}$, defined in terms of normalized mass, is the total thrust along the positive z-axis of the body frame.
  
  


The dynamic model of the $i$th UAV is defined as

\begin{subequations}\label{equ:dynamic}
	\begin{align}
	& \boldsymbol{\dot{p}}^i(t)=\boldsymbol{v}^i(t), \\
	& \boldsymbol{\dot{v}}^i(t)=\boldsymbol{R}^i\boldsymbol{l}_gT^i-\boldsymbol{l}_gg-\boldsymbol{D}_a^i\boldsymbol{v}^i(t), \\
	& \dot{\phi^i}(t)=\frac{1}{\tau_\phi^i}\left(K_\phi^i \phi_{\mathrm{ref}}^i(t)-\phi^i(t)\right), \\
	& \dot{\theta^i}(t)=\frac{1}{\tau_\theta^i}\left(K_\theta^i \theta_{\mathrm{ref}}^i(t)-\theta^i(t)\right),\\
    &\dot{\psi}^i(t) = \psi^i_{rate},
\end{align}
\end{subequations}
where $\boldsymbol{R}^i$ is the rotation matrix corresponding to the Euler angles, $\boldsymbol{D}_a^i = \mathrm{diag}(d_x^i, d_y^i, d_z^i)\in\mathbb{R}^{3 \times 3}$ is the damping coefficient matrix with $d_x^i, d_y^i$ and $d_z^i$ being positive constants, $g\in \mathbb{R}$ represents the gravitational acceleration, and $\boldsymbol{l}_g=[0,0,1]^{\mathrm{T}}$ denotes the unit vector in the direction of gravitational acceleration.   $\tau_\phi^i, \tau_\theta^i \in \mathbb{R}$ are the time constants of the $i$th UAV's first-order attitude dynamics, and $K_\phi^i, K_\theta^i \in \mathbb{R}$ are the corresponding gains.

 

We represent the dynamic model (\ref{equ:dynamic}) as $\boldsymbol{\dot{x}}^i=f(\boldsymbol{x}^i,\boldsymbol{u}^i)$.
To facilitate the design of NMPC, we discretize the dynamic model (\ref{equ:dynamic})
 using the forward Euler method at a step size of $h$:
\begin{equation}
	\boldsymbol{x}^i_{k+1} = \boldsymbol{x}^i_{k}+f(\boldsymbol{x}^i_{k},\boldsymbol{u}^i_{k})h,
	\label{closedloop}
\end{equation}
where $\boldsymbol{x}^i_k$ and $\boldsymbol{u}^i_k$ denote the state and input at the $k$th sampling time, respectively.
\subsection{Graph Theory for Formation Flying}
A directed graph $\mathcal{G}=(\mathcal{V}, \mathcal{E})$ is used to describe the network topology of the UAV formation, 
where $\mathcal{V}=\left\{\xi_1,\xi_2,...,\xi_n\right\}$ and $\mathcal{E}\subseteq\left\{(\xi_i,\xi_j),...,i \neq j,\xi_i, \xi_j \in \mathcal{V}\right\}$ are the node and edge set of the graph, respectively.
An edge $(i, j) \in \mathcal{E}$ indicates that agent $j$ is capable of receiving information from agent $i$, but not vice versa.
 Furthermore, node $\xi_i$ is the parent node, and node $\xi_j$ is the child node. We call node $\xi_i$ a neighbor of $\xi_j$.
The neighbor set of node $\xi_i$ is $\mathcal{N}_i$, and $\mathcal{N}_i=\left\{\xi_j \in \mathcal{V}:(\xi_i,\xi_j)\in\mathcal{E}\right\}$.

In a directed graph, a directed path from $\xi_i$ to $\xi_j$  is defined as a sequence of edges with the form 
$(\xi_i, \xi_1), (\xi_1, \xi_2), ..., (\xi_k, \xi_j)$. 
If every node, except for one, has exactly one parent, and there are directed paths from this parentless node to any other nodes, then we refer to the directed graph as a directed tree.
A directed spanning tree of a directed graph is a tree that encompasses all the graph's nodes. A directed graph possesses a directed spanning tree if the tree exists as a subset within the graph.

$\mathcal{A}=\left[a_{i j}\right] \in \mathbb{R}^{n \times \mathrm{n}}$ is used to represent the adjacency matrix associated with $\mathcal{G}$.
When node $\xi_i$ can obtain information of node $\xi_j$ , $a_{ij}>0$, otherwise, $a_{ij}=0$.
Self-edges are allowed in this paper.
We define the Laplacian matrix of $\mathcal{G}$ as $\mathcal{L}_{\mathcal{A}}=\left[l_{i j}\right] \in \mathbb{R}^{n \times n}$,
 and the Laplacian matrix can be obtained by $\mathcal{L}_{\mathcal{A}}=\mathcal{D}-\mathcal{A}$,
 where $\mathcal{D}=\operatorname{diag}\left(d_1, \ldots, d_n\right)$ is the degree matrix with $d_i=\sum_{j=1,j\neq i}^n a_{i j}$.
Hence, we can obtain $l_{i i}=d_i$, and $l_{i j}=-a_{i j}$, $i \neq j$.

\section{Control Algorithm Design}
\subsection{MRACon-Based Distributed NMPC Framework}
In this subsection, we propose a distributed control framework specifically for the UAV formation system.
Inspired by \cite{mei2021unified}, we utilize the model reference adaptive consensus (MRACon) framework to 
simplify the formation control algorithm designing process.
 The intrinsic idea of the MRACon framework is similar to that of model reference adaptive control (MRAC), 
 which involves designing a controller to track the 
 output of a linear reference model. By designing the tracking controller,
  the MRACon framework enables the actual system to achieve consensus.
 \par
\begin{figure}[!t]
	\centering
	\includegraphics[width=8.5cm]{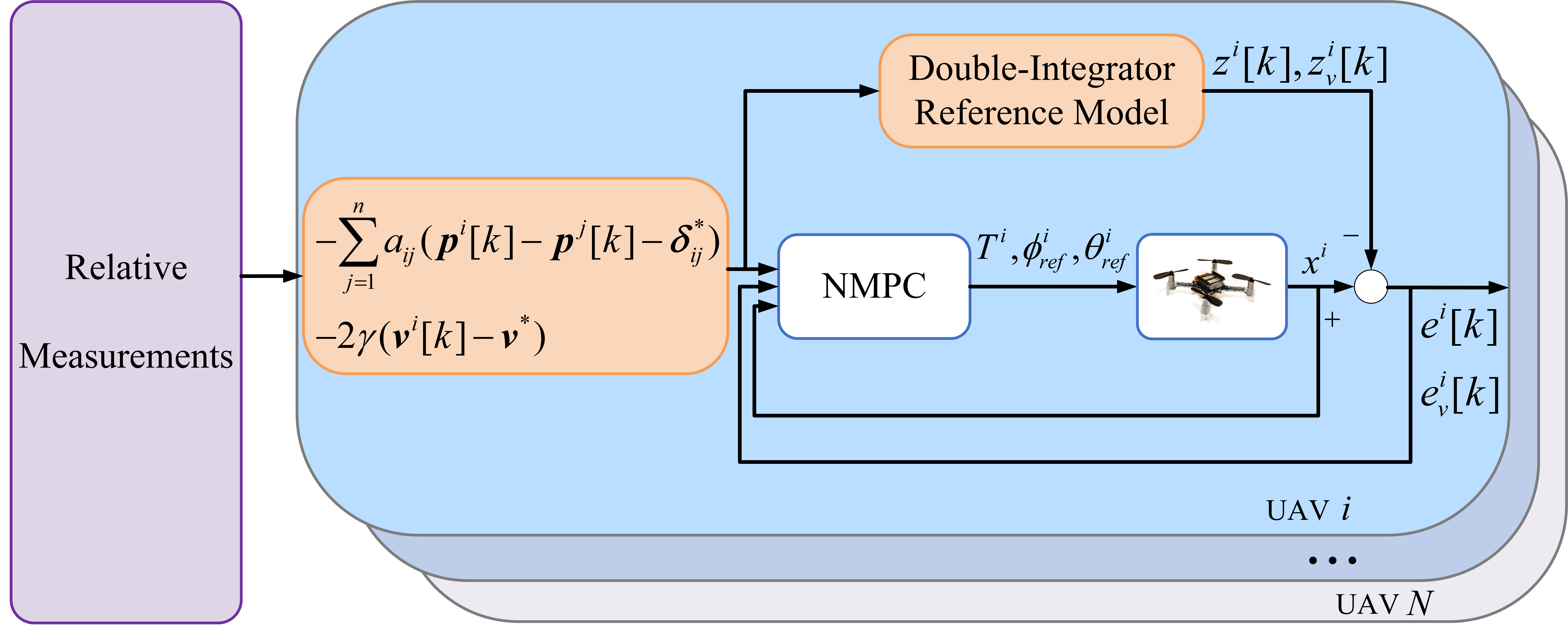}
	\caption{Block diagram of a multi-UAV formation flying system with the MRACon-based distributed NMPC algorithm.}
	\label{fig: MRACon-NMPC}
   \end{figure}

Due to the fact that the formation flying with collision avoidance problem has great complexity in the formation control, 
the nonlinear dynamics, and the safety requirements. We divide the problem 
into two subproblems: $(1)$ the formation control under the linear reference model;
 $(2)$ the collision-free tracking control under the UAV's nonlinear model.
A distributed formation algorithm is designed to calculate the output of the reference model. 
Due to the fact that NMPC is capable of handling complex constraints, especially collision avoidance constraints,
and is suitable for nonlinear dynamic systems. An NMPC algorithm is developed to track the output of the reference model.
Note that the formation algorithm for the reference model must accommodate discrete dynamics.
This requirement arises because nonlinear model predictive control (NMPC) tracks discrete desired states.
The MRACon-based distributed NMPC algorithm scheme is illustrated in \mbox{Fig. \ref{fig: MRACon-NMPC}}.
\subsection{Linear Reference Model Design}
In this subsection, we design a distributed formation algorithm for the linear reference model,
 modeling the UAV as a discrete double-integrator system. 
 By using the forward difference approximation from \cite{casbeer2008discrete}, the discrete dynamics of the reference model are shown as follows

   \begin{flalign}
	\begin{split}
	\boldsymbol{z}^{i}[k+1]&=\boldsymbol{z}^{i}[k]+h\boldsymbol{z_v}^i[k],\\
	\boldsymbol{z}_v^i[k+1]&=\boldsymbol{z}_v^i[k]+h\boldsymbol{u}^i_{f}[k],
	\label{referencemodel_closeloop}
   \end{split}
   \end{flalign}
   where $h$ is the step-size, $k$ is the discrete-time index, $\boldsymbol{z}^{i}[k]\in\mathbb{R}^3$ and $\boldsymbol{z}_v^{i}[k]\in\mathbb{R}^3$ are the reference model outputs,
   $\boldsymbol{u}^i_{f}[k]$ is the control input of the reference model.
   Inspired by a distributed consensus algorithm designed for discrete second-order dynamics under directed graphs \cite{cao2010multi}, we design a formation algorithm for UAV$i$:
\begin{flalign}
 \begin{split}
 \boldsymbol{u}^i_{f}[k]=\sum_{j=1}^n a_{i j}\left(\boldsymbol{p}^j[k]-\boldsymbol{p}^i[k]-\boldsymbol{\delta}_{i j}^*\right)
 -2{\gamma}\left(\boldsymbol{v}^i[k]-\boldsymbol{v}^*\right),
 \label{formation_second_order}
\end{split}
\end{flalign}
where $\gamma\in \mathbb{R}$ is a positive scalar,
$\boldsymbol{\delta}_{ij}^*=\boldsymbol{\delta}_i^*-\boldsymbol{\delta}_j^*$ is the desired relative position,
$\boldsymbol{\delta}_i^*\in \mathbb{R}^3$ is the desired absolute position, 
$\boldsymbol{v}^*\in \mathbb{R}^3$ is the prescribed formation maneuvering velocity,
$a_{ij}$ are the entries of the adjacency matrix $\mathcal{A}$, $i,j=1,...,n$.

Define the tracking errors between the reference model and the nonlinear dynamics model as
\begin{equation}
	 \boldsymbol{e}^i[k]=\boldsymbol{p}^i[k]-\boldsymbol{z}^i[k],\quad \boldsymbol{e}_v^i[k]=\boldsymbol{v}^i[k]-\boldsymbol{z}_v^i[k].
	 \label{trackingerr}
\end{equation}
According to (\ref{formation_second_order}), we can get
\begin{flalign}
	\begin{split}
	\boldsymbol{u}^i_{f}[k]=\sum_{j=1}^n a_{i j}\left(\boldsymbol{z}^j[k]-\boldsymbol{z}^i[k]-\boldsymbol{\delta}_{i j}^*\right)
	-2{\gamma}\left(\boldsymbol{z}_v^i[k]-\boldsymbol{v}^*\right)\\
	+\sum_{j=1}^n a_{i j}\left(\boldsymbol{e}^j[k]-\boldsymbol{e}^i[k]-\boldsymbol{\delta}_{i j}^*\right)
	-2{\gamma}\left(\boldsymbol{e}_v^i[k]-\boldsymbol{v}^*\right).
	\label{formation_second_order_err}
   \end{split}
   \end{flalign}
To perform the formation control analysis, we define:
\begin{equation*}
	\begin{split}
		\boldsymbol{\tilde{z}}^i[k]&=\boldsymbol{z}^i[k]-\boldsymbol{\delta_i^*}, \boldsymbol{\tilde{z}}_v^i[k]=\boldsymbol{z}_v^i[k]-\boldsymbol{v^*}, \\
		\boldsymbol{\tilde{y}}^i[k]&=\tilde{\boldsymbol{z}}^i[k]+\left(1 / \gamma\right) \tilde{\boldsymbol{z}}_v^i[k],
	\end{split}
	\end{equation*}
we then define the vector as
	\begin{equation*}
		\begin{split}
		\boldsymbol{\xi}[k]&=\left[\boldsymbol{\tilde{z}}^1[k], \boldsymbol{\tilde{y}}^1[k],\ldots, \boldsymbol{\tilde{z}}^n[k], \boldsymbol{\tilde{y}}^n[k]\right]^{\mathrm{T}}\in\mathbb{R}^{6n},\\
		\boldsymbol{\vartheta }[k]&=\left[\boldsymbol{e}^1[k], \boldsymbol{e}_v^1[k],\ldots, \boldsymbol{e}_v^n[k], \boldsymbol{e}_v^n[k]\right]^{\mathrm{T}}\in\mathbb{R}^{6n}.
	\end{split}
	\end{equation*}
	Then, by applying (\ref{formation_second_order}), we can obtain the dynamics of $\boldsymbol{\xi}[k]$:
	\begin{flalign}
		\begin{split}
			\boldsymbol{\xi}[k+1]&=\left[\left(\boldsymbol{I}_n \otimes \boldsymbol{A}-\boldsymbol{L}[k] \otimes \boldsymbol{B}\right) \otimes \boldsymbol{I}_3\right] \boldsymbol{\xi}[k]\\
			&+\left[\left(\boldsymbol{I}_n\otimes\boldsymbol{C}-\boldsymbol{L}[k] \otimes \boldsymbol{D}\right)\otimes\boldsymbol{I}_3\right]\boldsymbol{\vartheta}[k]\\
			&=\left(\boldsymbol{\Xi}[k] \otimes \boldsymbol{I}_3\right) \boldsymbol{\xi}[k]+\left(\boldsymbol{\Psi}[k]\otimes\boldsymbol{I}_3\right)\boldsymbol{\vartheta}[k],
		\label{referencemodel_discrete}
		\end{split}
	\end{flalign}
	where $\otimes$ denotes the Kronecker product operator; $\boldsymbol{L}[k]$ is the Laplacian matrix of the formation system;
	$\boldsymbol{\Xi}[k]=\boldsymbol{I}_n \otimes \boldsymbol{A}-\boldsymbol{L}[k] \otimes \boldsymbol{B}$;
	$\boldsymbol{\Psi}[k]=\boldsymbol{I}_n \otimes \boldsymbol{C}-\boldsymbol{L}[k] \otimes \boldsymbol{D}$;
	$\boldsymbol{A}$, $\boldsymbol{B}$, $\boldsymbol{C}$ and $\boldsymbol{D}$ are expressed as follows
	\begin{equation}
		\begin{aligned}
		\boldsymbol{A}&=\left[\begin{array}{cc}
		-\gamma h+1 & \gamma h \\
		\gamma h & -\gamma h+1
		\end{array}\right], \quad \boldsymbol{B}=\left[\begin{array}{ll}
		0 & 0 \\
		\frac{h}{\gamma} & 0
		\end{array}\right],\\
		\boldsymbol{C}&=\left[\begin{array}{cc}
			0 &  0 \\
			0 & -2h
			\end{array}\right], \quad \boldsymbol{D}=\left[\begin{array}{ll}
			0 & 0 \\
			\frac{h}{\gamma} & 0
			\end{array}\right].
		\end{aligned}
	\end{equation}
    If $\boldsymbol{e}^i[k]=\boldsymbol{0}_3$ and $\boldsymbol{e}^i_v[k]=\boldsymbol{0}_3$ satisfy, then $\boldsymbol{\vartheta}[k]=\boldsymbol{0}_{6n}$ holds. Therefore, (\ref{referencemodel_discrete}) can be regarded as a double-integrator dynamic system under (\ref{formation_second_order}).
And the following lemma is obtained.

\textit{Lemma 3.1}:\ (\cite{qin2012consensus}) Assume that the setup of $\gamma$ guarantees that $\boldsymbol{\Xi}[k]$ is nonnegative and that its diagonal elements are positive.
Then, $\tilde{\boldsymbol{\xi}}[k]$ can be regarded as a multi-agent system associated with a digraph $\overline{\mathcal{G}}$, which has $2n$ nodes. 
The digraph $\mathcal{G}$ has a spanning tree if and only if the digraph $\overline{\mathcal{G}}$ also has a spanning tree.

If a matrix $\boldsymbol{F}$ is non-negative and all of its rows sums are 1, $\boldsymbol{F}$ is called a stochastic matrix.

\textit{Lemma 3.2}:\ (\cite{horn2012matrix}) If $\boldsymbol{F}$ is a stochastic matrix whose eigenvalue $\lambda=1$  has an algebraic multiplicity 1,
 and all other eigenvalues satisfy $|\lambda|<1$, then $\lim _{m \rightarrow \infty} \boldsymbol{F}^m=\mathbf{1}_n \boldsymbol{w}^{\mathrm{T}}$, where
 $\boldsymbol{w}^{\mathrm{T}} \boldsymbol{F}=\boldsymbol{w}^{\mathrm{T}}$ and $\boldsymbol{w}^{\mathrm{T}} \mathbf{1}_n=1$.

 \textit{Lemma 3.3}:\ (\cite{ren2005consensus}) If $\boldsymbol{F}$ is a nonnegative matrix which has the same constant row sums $\mu>0$, then $\mu$ is an eigenvalue of $\boldsymbol{F}$ with eigenvector $\boldsymbol{1}_n$. Moreover, the eigenvalue $\mu$ of matrix $\boldsymbol{F}$
 has algebraic multiplicity $1$, if and only if the graph associated with $\boldsymbol{F}$ has a spanning tree. In addition, if the graph
 has a spanning tree and the diagonal elements of $\boldsymbol{F}$ are all positive, then $\mu$ is the unique eigenvalue of the maximum modulus.

 \textit{Assumption 3.1}$:$ The information interaction topology among the UAVs in the formation is a fixed directed graph with a directed spanning tree.


\textit{Theorem 3.1}$:$ For (\ref{referencemodel_discrete}) under Assumptions 3.1, with
the velocity damping gain satisfying $\sqrt{\Delta } \leq \gamma<\frac{1}{h}$, where $\Delta=max_i \left\{d_i\right\}$,
if the tracking errors $\boldsymbol{e}^i[k]=\boldsymbol{0}$ and $\boldsymbol{e}^i_v[k]=\boldsymbol{0}$, the system can asymptotically achieve the formation by using (\ref{formation_second_order}).
Specifically, $\lim _{k \rightarrow \infty}\left(\boldsymbol{z}_i[k]-\boldsymbol{z}_j[k]\right)=\boldsymbol{\delta}_{i j}^*$, $\lim _{k \rightarrow \infty} \boldsymbol{z}^v_i[k]=\boldsymbol{v}^* \ \forall i, j=1, \ldots, n$.

\begin{proof}
If the tracking errors are zero, then $\boldsymbol{\xi}[k+1]=\left(\boldsymbol{\Xi}[k] \otimes \boldsymbol{I}_3\right) \boldsymbol{\xi}[k]$.
The matrix $\boldsymbol{\Xi}\in\mathbb{R}^{2n}$ can be decomposed into an $n \times n$ block matrix:
\begin{equation}
	\left(\begin{array}{ccc}
	\boldsymbol{\Gamma}_{11}  & \cdots & \boldsymbol{\Gamma}_{1n} \\
	\vdots & \ddots &  \vdots \\
	\boldsymbol{\Gamma}_{n1}  & \cdots & \boldsymbol{\Gamma}_{nn} 
	\end{array}\right),
	\end{equation}
where the submatrix $\boldsymbol{\Gamma}_{ii}\in\mathbb{R}^2$ on the diagonal and the submatrix $\boldsymbol{\Gamma}_{ij}\in\mathbb{R}^2$ on the off-diagonal are expressed as
\begin{equation}
	\boldsymbol{\Gamma}_{ii}=\left[\begin{array}{cc}
	-\gamma h+1 & \gamma h \\
	\gamma h - \frac{d_i h}{\gamma} & -\gamma h+1
	\end{array}\right], \quad \boldsymbol{\Gamma}_{ij}=\left[\begin{array}{ll}
	0 & 0 \\
	\frac{a_{ij}h}{\gamma} & 0
	\end{array}\right],
\end{equation}
    We can derive that the elements on the main diagonal of the matrix $\boldsymbol{\Xi}[k]$ are positive if $\gamma<1/h$, 
	and the off-diagonal elements of the matrix $\boldsymbol{\Xi}[k]$ are non-negative if $\gamma \geq \sqrt{\Delta}$,
	where $\Delta$ is the maximum value of $d_i$.
Furthermore, the matrix $\boldsymbol{\Xi}$ satisfies
\begin{equation}
	\begin{aligned}
	\boldsymbol{\Xi} \mathbf{1}_{2 n}& =\left(\boldsymbol{I}_n \otimes \boldsymbol{A}-(\boldsymbol{L}[k] \otimes \boldsymbol{B}) \right)\left(\mathbf{1}_n \otimes \boldsymbol{1}_2 \right) \\
	& =\mathbf{1}_n \otimes\left(\boldsymbol{A} \boldsymbol{1}_2\right)-L[k] \mathbf{1}_n \otimes\left(\boldsymbol{B} \boldsymbol{1}_2\right)\\
	& =\mathbf{1}_{2 n}-\mathbf{0}_{2 n} =\mathbf{1}_{2 n}.
	\end{aligned}
	\end{equation}
	Thus, $\boldsymbol{\Xi}$ is a stochastic matrix with positive diagonal elements.
	Based on Lemma $3.1$, if $\mathcal{G}$ has a spanning tree, then $\overline{\mathcal{G}}$ also has a spanning tree.
	According to Lemma $3.3$, the algebraic multiplicity of $\lambda=1$ is 1, and the remaining eigenvalues satisfy $| \lambda|<1$.
	Furthermore, from Lemma $3.2$, it follows that $\lim _{m \rightarrow \infty} \boldsymbol{\Xi}^m=\mathbf{1}_n$.
	Thus, we can obtain
	\begin{equation}
		\begin{aligned}
		\lim _{k \rightarrow \infty} \tilde{\boldsymbol{\xi}}[k] & =\lim _{k \rightarrow \infty}\left(\boldsymbol{\Xi}^k \otimes \boldsymbol{I}_p\right) \boldsymbol{\xi}[0] \\
		& =\left(\mathbf{1}_{2 n} \boldsymbol{w}_1^{\mathrm{T}} \otimes \boldsymbol{I}_p\right) \boldsymbol{\xi}[0]\\
		& =\left(\mathbf{1}_{2 n} \otimes \boldsymbol{I}_p\right)\left(\boldsymbol{w}_1^{\mathrm{T}} \otimes \boldsymbol{I}_p\right) \boldsymbol{\xi}[0] \\
		& = \mathbf{1}_{2 n} \otimes \boldsymbol{\zeta},
		\end{aligned}
		\end{equation}
where $\boldsymbol{\zeta}\in \mathbb{R}^3$ is a constant vector
defined as $\boldsymbol{\zeta}=\left(\boldsymbol{w}_1^{\mathrm{T}} \otimes \boldsymbol{I}_p\right) \boldsymbol{\xi}[0]$.
Therefore, we have $\tilde{\boldsymbol{z}}^i[k] \rightarrow \tilde{\boldsymbol{z}}^j[k]$ as $k \rightarrow \infty$.
Consequently, $\tilde{\boldsymbol{z}}^i_v[k]=\gamma\left(\tilde{\boldsymbol{y}}^i[k]-\tilde{\boldsymbol{z}}^i[k]\right)\rightarrow0$ as $k \rightarrow \infty$.
This indicates that $\lim _{k \rightarrow \infty}\left(\boldsymbol{z}^i[k]-\boldsymbol{z}^j[k]\right)=\boldsymbol{\delta}_{i j}^*$, $\lim _{k \rightarrow \infty} \boldsymbol{z}_v^i[k]=\boldsymbol{v}^*$.
 In other words, the system can asymptotically achieve the prescribed formation under (\ref{formation_second_order}) when tracking errors $\boldsymbol{e}^i$ and $\boldsymbol{e}_v^i$ are zero.
\end{proof}

\subsection{NMPC Tracking Algorithm Design}
In this subsection, we design an NMPC controller over a horizon length $N\in \mathbb{N}^{+}$
to track the output of the reference model.
 In what follows, the notation $\hat{(\cdot)}\left[l \mid k\right]$ represent the predicted value at $k+l$ 
 with the information acquired at $k$, where $l\in\left\{0,\dots,N\right\}$.
 First, we formulate the NMPC cost function 
to minimize the tracking error and the control effort.

$1)$ \emph{Tracking Error Penalty}:
\begin{equation}
 \ell_{e}^i=\left\|\hat{\boldsymbol{x}}^i[l|k]-\boldsymbol{x}_{r}^i[l|k]\right\|_2^{\boldsymbol{Q}^i},
\end{equation}
where $\boldsymbol{Q}^i_x\in\mathbb{R}^{9 \times 9}$ is the weighting matrix, 
$\hat{\boldsymbol{x}}^i[l|k]\in\mathbb{R}^9$ and $\boldsymbol{x}_{r}^i[l|k]=[\boldsymbol{z}^i[k+l], \boldsymbol{z}_v^i[k+l], \overline{\phi}^i_r,\overline{\theta}^i_r,\overline{\psi}^i_r]^{\mathrm{T}}\in\mathbb{R}^9$ are the state and the reference state of the $i$th UAV. 
$\boldsymbol{z}^i[k+l]$ and $\boldsymbol{z}^i_v[k+l]$ are the outputs of the reference model at $k+l$.
 The terms $\overline{\phi}^i_r,\overline{\theta}^i_r$ denote the reference of the roll and pitch angles derived from the reference model.
 These angles can be determined by the differential flatness property.
 $\psi^i$ is the reference of the yaw angle.
According to (\ref{referencemodel_closeloop}), the dynamics of the linear reference model at $k+l$ are expressed as

\begin{flalign}
	\begin{split}
	\boldsymbol{z}^{i}[k+l+1]&=\boldsymbol{z}^{i}[k+l]+h\boldsymbol{z_v}^i[k+l],\\
	\boldsymbol{z}_v^i[k+l+1]&=\boldsymbol{z}_v^i[k+l]+h\boldsymbol{u}^i_{f}[k+l].
	\label{referencemodel_closeloop_kl}
   \end{split}
   \end{flalign}
   Note that the output of the reference model at the future time steps $k+l (l\geq1)$ requires relative measurements in the future, which challenges the controller design.
   We use the current relative measurements to replace the corresponding measurements at the future time step, i.e., $\boldsymbol{u}^i_{f}[k+l]=\boldsymbol{u}^i_{f}[k]$,
 rather than communicate the predicted trajectory among the UAVs.
   Since NMPC applies only the first element of the control input sequence to the UAV, our approach is appropriate for this scenario.
   
$2)$ \emph{Control Effort Penalty}:
\begin{equation}
	\ell_{u}^i=\left\|\hat{\boldsymbol{u}}^i[l|k]-\boldsymbol{u}^i_{r}[l|k]\right\|_2^{\boldsymbol{Q}_u^i},
\end{equation}
where $\boldsymbol{Q}_u^i\in\mathbb{R}^{4\times 4}$ is the weighting matrix,
$\hat{\boldsymbol{u}}^i[l|k]\in\mathbb{R}^4$ and $ \boldsymbol{u}^i_{r}[l|k]=[g,\overline{\phi}^i_r,\overline{\theta}^i_r, 0]^{\mathrm{T}}\in\mathbb{R}^4$
 are the control input and the reference control input of the $i$th UAV.

Then, we design the constraint sets for NMPC, including 
 obstacle avoidance, reciprocal collision avoidance,
 and physical constraints.

 $1)$ \emph{Obstacle Avoidance Constraints}:
The constraints are established to prevent UAVs from colliding with static obstacles.
Assume that there are $M$ obstacles in the UAV's perception range, labeled as $m \in \mathcal{O}=\{1,2, \ldots, M\}$, 
where $m$ denotes the index of an obstacle, and $\mathcal{O}$ represents the set of all obstacles.
We assume that all obstacles can be enclosed by a set of cylinders, where the position of each cylinder's center is defined as
$\boldsymbol{p}_o^m=[x_o^m, y_o^m]^{\mathrm{T}}\in\mathbb{R}^2,m\in\mathcal{O}$. 
The constraints are shown as:
\begin{equation}
 \left\| \boldsymbol{L}_o\hat{\boldsymbol{p}}^i[l|k]-\boldsymbol{p}_o^m\right\|_2 \geq r_o^m, m \in \mathcal{O},
 \label{collision_avoidance}
\end{equation}
where $\boldsymbol{L}_o=\mathrm{diag}(1,1,0)\in\mathbb{R}^{3 \times 3}$, and $r_o^m>0$ represents the
 minimum safety distance of the cylindrical region. 

$2)$ \emph{Reciprocal Collision Avoidance Constraints}:
The reciprocal collision avoidance constraints are imposed to prevent UAVs from colliding with each other.
Especially, the high-speed rotating propellers of UAVs generate aerodynamic interactions among them, known as the downwash effect.
 Considering this, we build the collision model between the $i$th UAV and neighbor $j$ as an ellipsoid:
 \begin{equation}
	 \boldsymbol{C}^{ij}=\left\{\boldsymbol{s}\in\mathbb{R}^3|\left\|\boldsymbol{\Theta}^{-1}\boldsymbol{s}\right\|_2\leq1\right\},
 \end{equation}
where $\boldsymbol{\Theta} = \mathrm{diag}(\theta_a,\theta_b,\theta_c)\in\mathbb{R}^{3 \times 3}$, $\theta_a=\theta_b<\theta_c$.
A larger $\theta_c$ is designed to alleviate the downwash effect. 
We utilize a motion prediction with a collision-check mechanism to 
deal with the reciprocal avoidance problem.

\begin{figure}[!t]
	\centering
	\includegraphics[width=8.0cm]{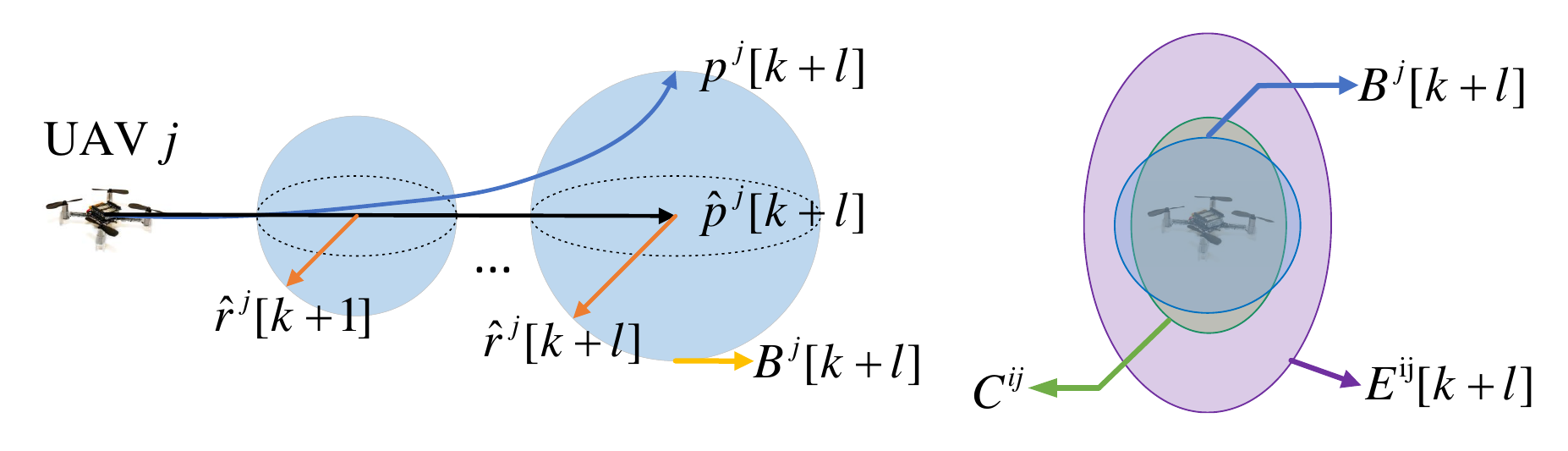}
	\caption{(Left) The $j$th neighboring UAV's predicted trajectory $\hat{\boldsymbol{p}}^j[k+l]$ and spherical error boundary $\boldsymbol{B}^j[k+l]$.
	 (Right) Collision model $\boldsymbol{C}^{ij}$, spherical error boundary $\boldsymbol{B}^j[k+l]$, and outer ellipsoid $\boldsymbol{E}^{ij}[k+l]$.}
	\label{fig:intercollision}
   \end{figure}
First, we build the neighbor's motion prediction model:
\begin{equation}
  \hat{\boldsymbol{p}}^j[k+l]=\boldsymbol{p}^j\left[k\right]+\boldsymbol{v}^j\left[k\right]lh,
\end{equation}
where $\hat{\boldsymbol{p}}^j[k+l]$ denotes the predicted trajectory of the $j$th UAV, $\boldsymbol{p}^j[k]$ and $\boldsymbol{v}^j[k]$ denote the current position and velocity of neighbors, and $h$ is the prediction step of NMPC.
Furthermore, considering that the constant velocity in the motion prediction model is not accurate enough, a time-variant spherical error boundary is introduced to describe the trajectory prediction error of the neighbors \cite{park2020online}, as shown in \mbox{Fig. \ref{fig:intercollision}}:
 \begin{equation}
	\boldsymbol{B}^j[k+l]=\left\{\boldsymbol{s}\in\mathbb{R}^3|\left\| \boldsymbol{s}\right\|_2 \leq \hat{r}^j[k+l]\right\},
 \end{equation}
where $ \hat{r}^j[k+l]$ is the time-variant radius of $\boldsymbol{B}^j[k+l]$:
\begin{equation}
 \hat{r}^j[k+l]=\frac{1}{2} T_{\max }^j \min \left(lh, M_rh\right)^2,
\end{equation}
where $T_{max}^j$ is the maximum acceleration of the $j$th UAV, and $M_r$ is a scalar used to limit $\hat{r}^j[k+l]$ in case the constraint is too conservative.
As depicted in \mbox{Fig. \ref{fig:intercollision}}, the reachable area of $\boldsymbol{p}^j[k+l]$ is expressed as
\begin{equation}
 \boldsymbol{p}^j[k+l] \in \left(\hat{\boldsymbol{p}}^j[k+l] \oplus \boldsymbol{B}^j[k+l]\right),
\end{equation}
where $\oplus$ is the Minkowski sum, $\boldsymbol{p}^j[k+l]$ is the actual trajectory of UAV$j$ at $k+l$.
 Then, the collision avoidance
 region of neighbors is expressed as $\boldsymbol{B}^j[k+l] \oplus \boldsymbol{C}^{ij}$.
 According to \cite{kurzhanski1997ellipsoidal}, 
an optimal outer ellipsoid $\boldsymbol{E}^{ij}[k+l]$ that contains the Minkowski sum $\boldsymbol{B}^j[k+l] \oplus \boldsymbol{C}^{ij}$ is shown as 
  \begin{equation}
  	\boldsymbol{E}^{ij}[k+l]  = (1+\beta)\boldsymbol{B}^j[k+l] + (1+\frac{1}{\beta})\boldsymbol{C}^{ij},
	\label{augmented_model}
  \end{equation} 
where $\beta=\sqrt{\frac{\text{tr}(\boldsymbol{B}^j[k+l])}{\text{tr}(\boldsymbol{C}^{ij})}}$, and the outer ellipsoid is denoted as
  \begin{flalign}
 \begin{split}
	{\boldsymbol{E}}^{ij}[k+l]= \left\{\boldsymbol{s}\in\mathbb{R}^3\mid \left\|\tilde{\boldsymbol{\Theta}}^{-1} \boldsymbol{s}\right\|_2 \leq 1\right\},
 \end{split}
\end{flalign} 
where $\tilde{\boldsymbol{\Theta}} = \mathrm{diag}(\tilde{\theta}_a,\tilde{\theta}_b,\tilde{\theta}_c)\in\mathbb{R}^{3 \times 3}$, $\tilde{\theta}_a=\tilde{\theta}_b<\tilde{\theta}_c$.

   Second, we utilize a collision-check mechanism to determine where the predicted trajectory of neighboring UAVs will result in a collision.
   The optimal predicted trajectory of the $i$th UAV, as determined in the previous horizon (i.e., $k-1$) of NMPC, in conjunction with the predicted motion of neighboring UAVs to conduct the collision check.
   Only the first collision that occurs with $j$th neighboring UAVs is considered, and the reciprocal constraints are built as
\begin{equation}
 \left\|\tilde{\boldsymbol{\Theta}}^{-1}\left(\hat{\boldsymbol{p}}^i[l_c|k]-\hat{\boldsymbol{p}}^j[k+l_c]\right)\right\|_2 \geq 1,
 \label{ondemand_ca}
\end{equation}
where $l_c$ denotes the moment when the $i$th UAV first detects a collision with the $j$th UAV.

The benefits of our approach are twofold:
$(1)$ the motion prediction mechanism avoids trajectory information communication among the UAVs;
$(2)$ the collision-check avoidance mechanism decreases the
 number of constraints, reducing the computational burden of NMPC.

$3)$ \emph{Physical Constraints}:
The UAVs' movements should conform to their physical characteristics.
Several physical constraints are considered in this paper, including flying area limitations, velocity limitations, and control input restrictions.
First, the UAVs are required to stay within a specified space, such as an indoor flying area:
\begin{equation}
	\boldsymbol{p}^i_{\min}\leq\hat{\boldsymbol{p}}^i[l|k]\leq \boldsymbol{p}^i_{\max},
	\label{physical_p}
   \end{equation}
where $\boldsymbol{p}^i_{min}, \boldsymbol{p}^i_{max}$ denote the boundaries of the flying area.
Second, the UAVs have a limited velocity upper bound:
\begin{equation}
	\left\|\hat{\boldsymbol{v}}^i[l|k]\right\|_2\leq v_{\max},
	\label{physical_v}
\end{equation}
where $v_{\max}$ is a positive constant.
Third, the UAV's rotors have limited actuation, and the reference angle can only vary
 within a specified range. Consequently, the control input constraints are defined as
\begin{equation}
 \left(\begin{array}{c}
 T_{\min } \\
 -\phi_{\max } \\
 -\theta_{\max } \\
 -\dot{\psi}_{\max}
 \end{array}\right) \leq \hat{\boldsymbol{u}}^i[l|k] \leq\left(\begin{array}{l}
 T_{\max } \\
 \phi_{\max } \\
 \theta_{\max }\\
 \dot{\psi}_{\max}
 \end{array}\right),
 \label{physical_a}
\end{equation}
where $0<T_{\min}<T_{\max}$, $\phi_{\max }$, $\theta_{\max }$ and $\dot{\psi}_{\max}$ are the positive constants.

The stage cost of the distributed NMPC is shown as
\begin{equation}
	\ell(\hat{\boldsymbol{x}}^i[l|k],\hat{\boldsymbol{u}}^i[l|k]) = \ell_{u}^i+\ell_{e}^i,
\end{equation} 
Accordingly, the optimal control problem for UAV$i$ is formulated as
\begin{flalign}
 \begin{split}
 &\boldsymbol{u}_{\mathrm{NMPC}}^i=\operatorname{argmin}_{\boldsymbol{u}} \sum^{N}_{l=0} \ell(\hat{\boldsymbol{x}}^i[l|k],\hat{\boldsymbol{u}}^i[l|k]),\\
 \text { s.t. }&\hat{\boldsymbol{x}}^i[l+1|k]=\hat{\boldsymbol{x}}^i[l|k]+f\left(\hat{\boldsymbol{x}}^i[l|k], \hat{\boldsymbol{u}}^i[l|k]\right)h,\\
 &\hat{\boldsymbol{x}}^i[l|k]=\boldsymbol{x}^i[k], \text{for} \ l=0,\\
 &\text{Constraints (\ref{collision_avoidance})}, \text{Constraints (\ref{ondemand_ca})},\\
 &\text{Constraints (\ref{physical_p}-\ref{physical_a})},\\
 &\hat{\boldsymbol{x}}^i[l|k]\in\mathbb{X}, \hat{\boldsymbol{u}}^i[l|k]\in\mathbb{U},
 \label{OCP}
 \end{split}
\end{flalign}
where $\mathbb{X},\mathbb{U}$ are the set of admissible states and inputs.
We can get the optimal control sequence $\boldsymbol{u}_{\mathrm{NMPC}}^i$ with $N$ inputs by solving (\ref{OCP}), but only the first element of $\boldsymbol{u}_{\mathrm{NMPC}}^i$ is
applied to UAVs.
The stability of NMPC without terminal state constraints has been well studied in \cite{grune2017nonlinear},
and we can conclude as follows.
The minimum stage cost is defined as
\begin{equation}
 \ell^{*}(\hat{\boldsymbol{x}}^i[l|k])=\inf _{\boldsymbol{u}\in\mathbb{U}} \ell(\hat{\boldsymbol{x}}^i[l|k], \hat{\boldsymbol{u}}^i[l|k]),
\end{equation}

\textit{Assumption 3.2}$:$ The minimum stage cost satisfies
\begin{align}
 &\nonumber\alpha_1\left(\left\|\hat{\boldsymbol{x}}^i[l|k]-\boldsymbol{x}_r^i[l|k]\right\|_2\right) \leqslant \ell^*( \hat{\boldsymbol{x}}^i[l|k]), \\
 &\ell^*( \hat{\boldsymbol{x}}^i[l|k]) \leqslant \alpha_2\left(\left\|\hat{\boldsymbol{x}}^i[l|k]-\boldsymbol{x}_r^i[l|k]\right\|_2\right),
\end{align}
where $\left\|\hat{\boldsymbol{x}}^i[l|k]-\boldsymbol{x}_r^i[l|k]\right\|_{2}$ denotes the norm of
the tracking error at the minimum stage cost; $\alpha_1, \alpha_2 \in K_{\infty}$, and $\hat{\boldsymbol{x}}^i[l|k] \in \mathbb{X}$.

\textit{Assumption 3.3}$:$ For all $\hat{\boldsymbol{x}}^i[l|k]\in \mathbb{X}$, there exists a feasible $\hat{\boldsymbol{u}}^i[l|k]\in\mathbb{U}$,
a scalar $C\geq1$ and $\sigma \in(0,1)$ satisfying
\begin{equation}
 \ell(\hat{\boldsymbol{x}}^i[l|k],\hat{\boldsymbol{u}}^i[l|k]) \leq C\sigma \ell^{i*}(\hat{\boldsymbol{x}}^i[l|k]).
\end{equation}

\textit{Lemma 3.4} \cite{grune2017nonlinear}$:$ Consider the NMPC problem with a prediction horizon $N\in\mathbb{N}^+$
under Assumptions 3.2 and 3.3, the closed-loop dynamic system is uniformly asymptotically stable on $\mathbb{X}$ provided that $N$ is sufficiently large.


Consequently, the main result of formation flying with collision avoidance is introduced as follows.

\textit{Theorem 3.2}$:$ 
A group of UAVs can achieve formation flying with no collision occurring by solving the OCP (\ref{OCP})
 provided that $N$ is sufficiently large and under Assumptions 3.1-3.3.

 \begin{proof}
	According to Lemma 3.4, the NMPC tracking algorithm ensures that 
	 $\lim _{k \rightarrow \infty} \boldsymbol{e}^i[k]=\mathbf{0}_3$, $\lim _{k \rightarrow \infty} \boldsymbol{e}_v^i[k]=\mathbf{0}_3$ holds.
	Then, we can get $\boldsymbol{p}^i[k]=\boldsymbol{z}^i[k]$, $\boldsymbol{v}^i[k]=\boldsymbol{z}^i_v[k]$ when $k\rightarrow \infty$.
	Furthermore, from Theorem 3.1, we can get that the reference model can achieve $\lim _{k \rightarrow \infty}\left(\boldsymbol{z}^i[k]-\boldsymbol{z}^j[k]\right)=\boldsymbol{\delta}_{i j}^*$, $\lim _{k \rightarrow \infty} \boldsymbol{\dot{z}}^i_v[k]=\boldsymbol{v}^*$ by using (\ref{formation_second_order}).
	Based on the above analysis, we can conclude that $\lim _{k \rightarrow \infty}\left(\boldsymbol{p}^i[k]-\boldsymbol{p}^j[k]\right)=\boldsymbol{\delta}_{i j}^*$, $\lim _{k \rightarrow \infty} \boldsymbol{v}^i[k]=\boldsymbol{v}^*$, the UAVs form a formation and reach the desired formation flight velocity.
	Hard constraints of obstacle avoidance and inter-collision avoidance in the OCP can ensure flight safety.
	Therefore, UAVs can achieve formation flight and collision avoidance.
 \end{proof}

 \section{Simulation and hardware experiments}
 \subsection{Simulation}
 In this subsection, we conducted several simulation experiments to show 
 the algorithm's characteristics.
 Initially, we develop a fixed formation simulation to evaluate the safety and efficiency of the UAV swarm navigation through a complex environment. Subsequently, we design a simulation with formation transformations to further validate the algorithm further. The parameters utilized in the simulation are presented in \mbox{Table \ref{Table_1}}.

\begin{figure}[thpb]
	\centering
	\includegraphics[width=9.0cm]{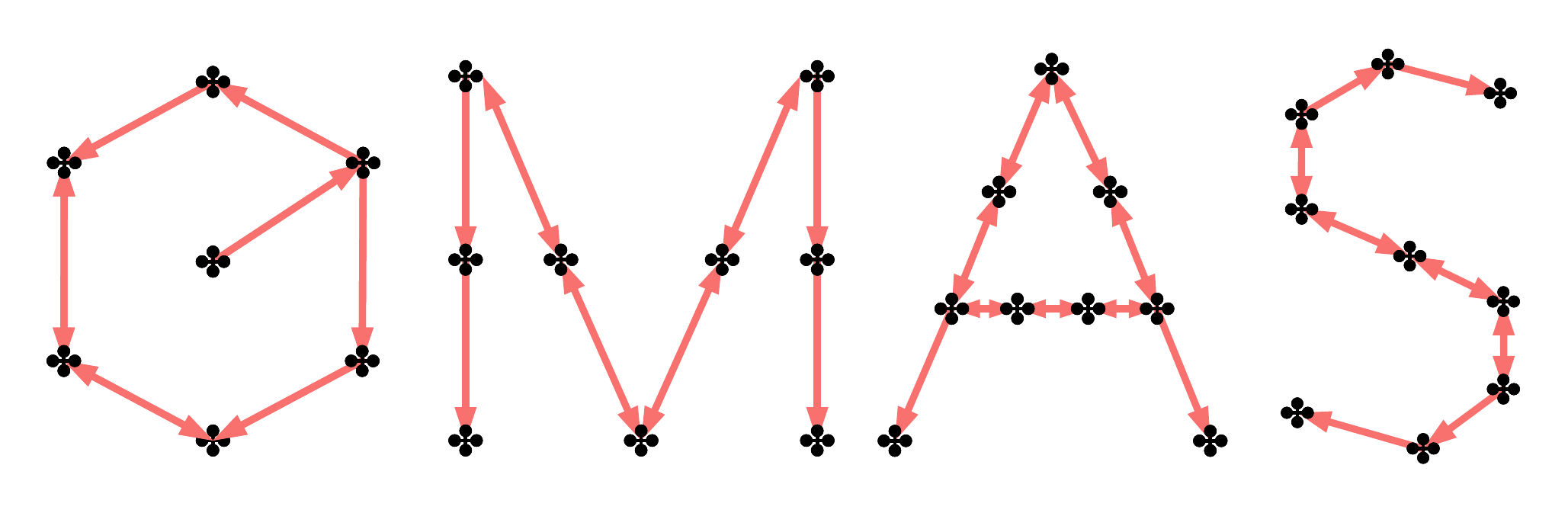}
	\caption{Directed communication topologies with corresponding geometric shapes, from left to right:
	a hexagon used in the fixed formation experiment;
	followed by the letters ``M", ``A", and ``S" used in the formation transformation experiment.}
    \label{fig:Topology}
\end{figure}

  \begin{table}[!t]
	\renewcommand{\arraystretch}{1.6}
	\caption{SIMULATION PARAMETER}
	\centering
	\label{table_1}
	\resizebox{\columnwidth}{!}{
		\begin{tabular}{|l|c|}
			\hline Parameter & Value \\
			\hline$[\tau_\phi^i,\tau_\theta^i]$ & {$[0.116,0.116]$} \\
			\hline$[K_\phi^i,K_\theta^i]$ & {$[1,1]$} \\
			\hline$[d_x^i, d_y^i, d_z^i]$ & {$[0.1,0.1,0.2]$} \\
			\hline$g$ & {$9.81[m/s^2]$} \\
			\hline$\left[T_{\min}, -\phi_{\max}, -\theta_{\max}, -\dot{\psi}_{\max} \right]$ & {$[5.0,-0.4,-0.4, -0.1]$} \\
			\hline$\left[T_{\max}, \phi_{\max}, \theta_{\max},\dot{\psi}_{\max} \right]$ & {$[15.0,0.4,0.4, 0.1]$} \\
			\hline$\gamma$ & 2 \\
			\hline$M_r$ & 2 \\
			\hline$\boldsymbol{Q}^i$ & {$100*\text{diag}(1,1,3,1,1,1,0.1,0.1,0.1)$} \\
			\hline$\boldsymbol{Q}^i_u$ & {$\text{diag}(1,1,1,1)$} \\
			\hline$\left[\theta_a,\theta_b,\theta_c \right]$ & {$[0.6,0.6,0.72]$} \\
			\hline Sample period $h$ & 0.05s \\
			\hline Prediction horizon $N$ & 10 \\
			\hline
			\end{tabular}
	}

	\label{Table_1}
\end{table}
Four directed communication topologies with corresponding geometric shapes are defined in \mbox{Fig. \ref{fig:Topology}}.
 An arrow pointing from UAV $i$ to UAV $j$ indicates that UAV $j$ can acquire 
information from UAV $i$. Conversely, if there is no arrow between UAV $i$ and UAV $j$, there is no information interaction between them.

\begin{figure}[thpb]
	\centering
	\includegraphics[width=9.0cm]{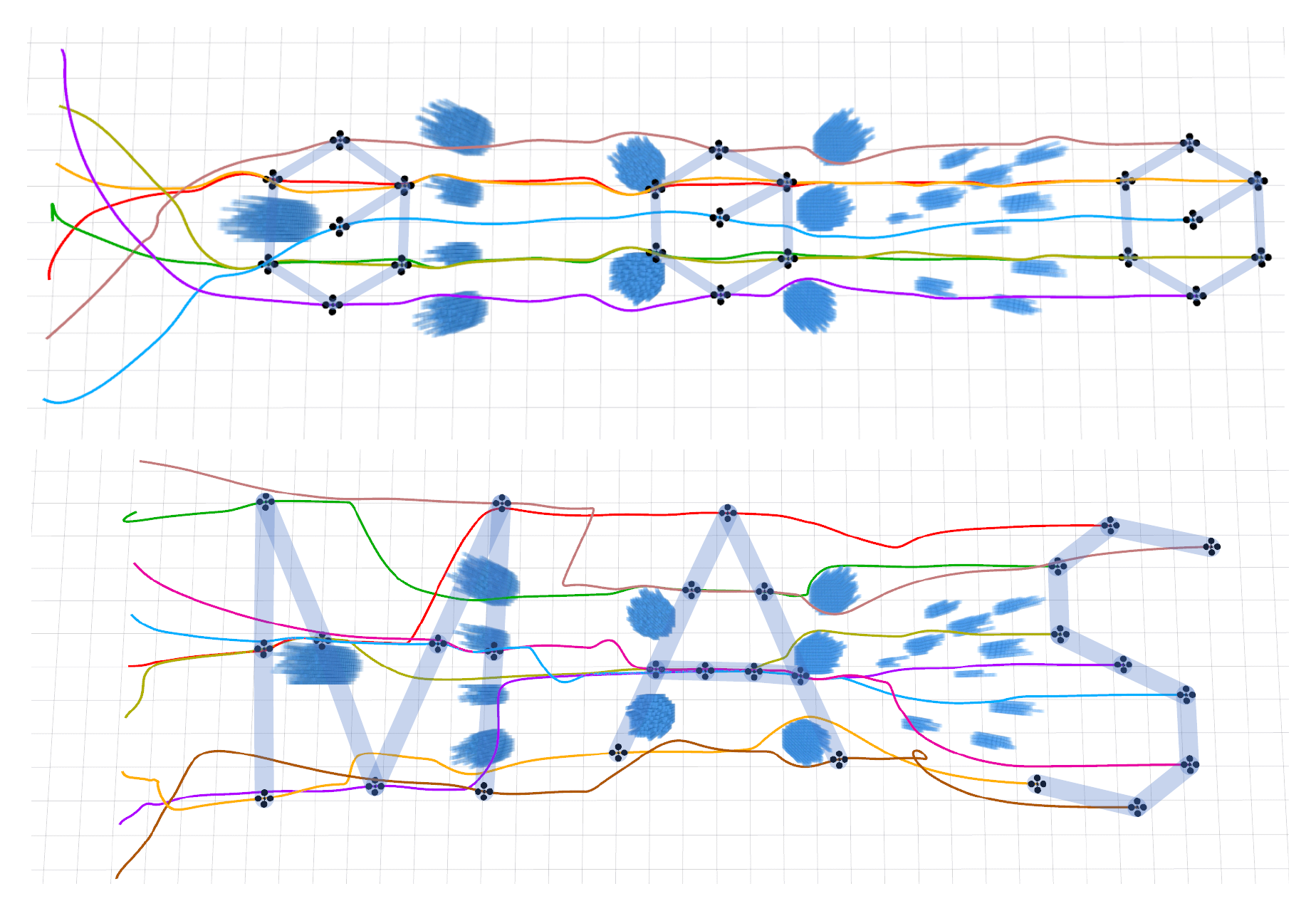}
	\caption{Formation trajectories of the UAVs flying from the left side to the right side. The black symmetrical shapes represent the UAVs, 
	the blue columns represent obstacles, and the thick light blue lines represent the edges of the UAV formation.
	The upper graph displays the result of a hexagon-shaped UAV formation flying through a cluttered environment. 
	The below shows the result of UAV formation transformation during the flight through the environment with formation shapes like the letters ``M", ``A", and ``S".}
    \label{fig:UAV_Traj}
\end{figure}

\begin{figure}[thpb]
	\centering
	\includegraphics[width=9.0cm]{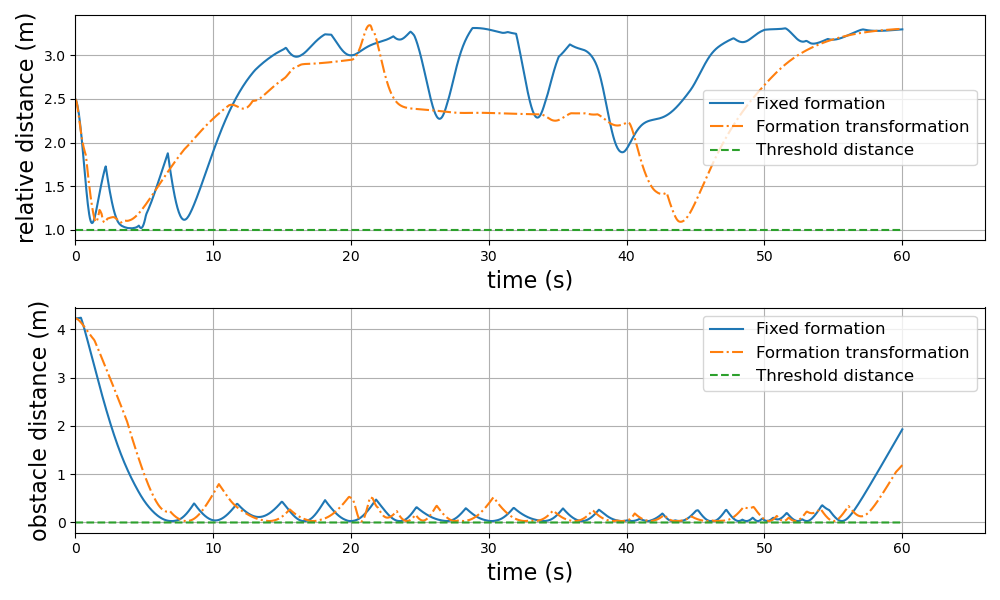}
	\caption{The upper graph shows the minimum relative distance among all UAVs, and the below shows the minimum distance between all UAVs and the surface of obstacles.}
    \label{fig:safe_distance}
\end{figure}

\begin{figure}[htbp]
	\centering
	\includegraphics[width=9.0cm]{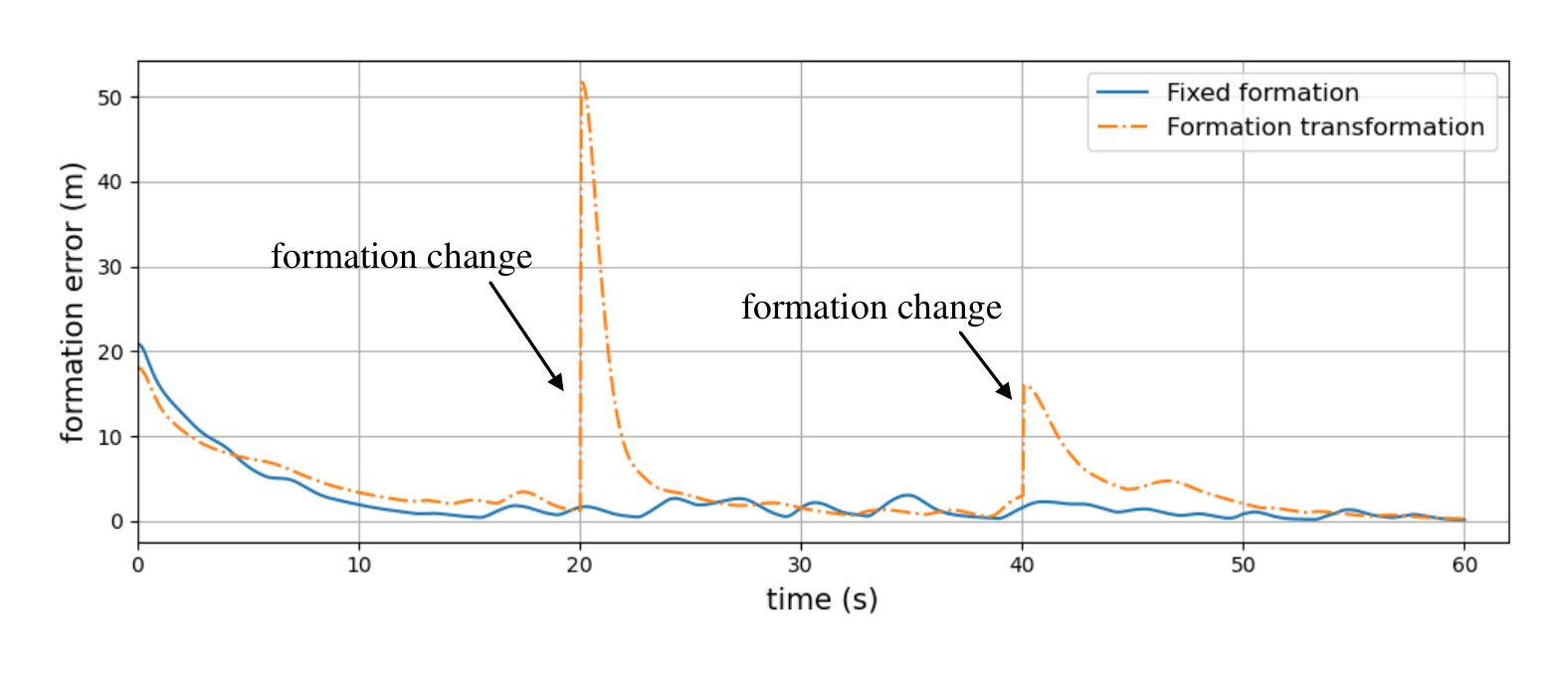}
	\caption{Formation error $\varepsilon_f$.}
    \label{fig:formation_error}
\end{figure}

\begin{figure}[htbp]
	\centering
	\includegraphics[width=9.0cm]{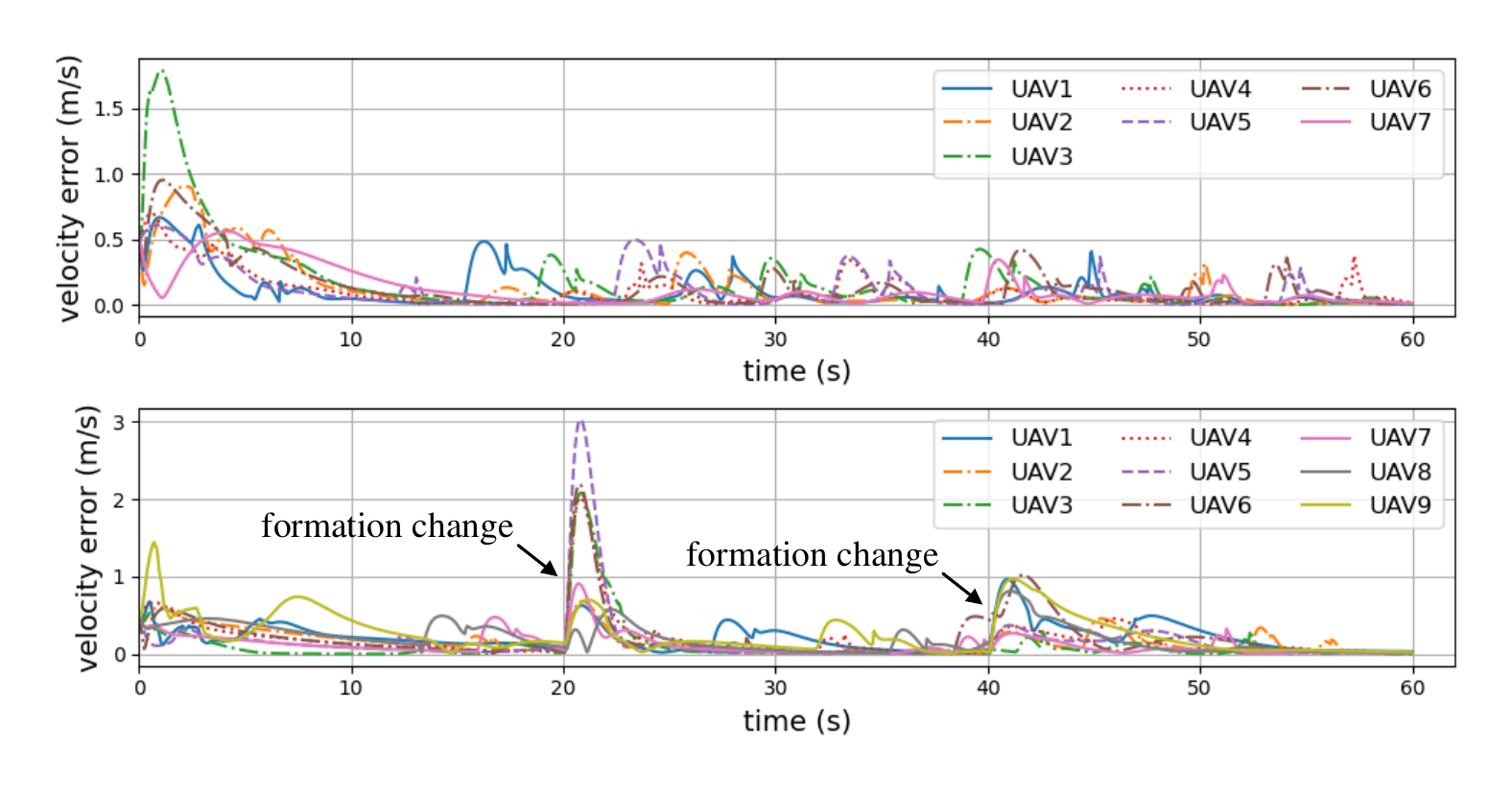}
	\caption{Velocity tracking error $\varepsilon_v^i$ of all UAVs.}
    \label{fig:velocity_tracking_error}
\end{figure}

The simulation results are shown in \mbox{Figs. \ref{fig:UAV_Traj}-\ref{fig:velocity_tracking_error}}.
\mbox{Fig. \ref{fig:UAV_Traj}} shows the formation trajectories.
In the upper graph, seven UAVs fly through a cluttered environment in a fixed, hexagon-shaped formation.
The initial positions of the UAVs in $x$-axis are set to 0 m, in $z$-axis are set to 1.5 m,
 and in $y$-axis are set to [4.5,3.0,1.5,0,-1.5,-3.0,-4.5] m, corresponding to UAV 1$-$7.
To further demonstrate the algorithm's capability, we conduct a simulation experiment on a group of nine UAVs
flying through a cluttered environment with three types of formation configuration with the shape of the letters ``M", ``A", and ``S",
 as shown in the lower graph of \mbox{Fig. \ref{fig:UAV_Traj}}.
The initial positions of the UAVs in $x$-axis are set to 0 m, in $z$-axis are set to 1.5 m,
 and in $y$-axis are set to [6.0,4.5,3.0,1.5,0,-1.5,-3.0,-4.5,-6.0] m, corresponding to UAVs 1$-$9.

\mbox{Fig. \ref{fig:safe_distance}} shows the minimum relative distance among all UAVs, as well as the minimum distance between the UAVs and the surfaces of the obstacle cylinders.
The relative distance is defined as $d_{ij} = \left\|\boldsymbol{\Theta}^{-1}(\boldsymbol{p}_i-\boldsymbol{p}_j)\right\|_2$,
and the threshold is set to 1. When $d_{ij} > 1$, there are no collision risks among any UAVs.
The obstacle distance is $d_{im} = \Vert \boldsymbol{p}^i\boldsymbol{L}_o-\boldsymbol{p}_o^m\Vert_2 - r_o^m$
and the threshold is set to 0. When $d_{im} > 0$ is satisfied, there are no collision risks between UAVs and obstacles.
From the results, it can be seen that MRACon-based NMPC can ensure the safety of UAVs during the formation flight.

\mbox{Fig. \ref{fig:formation_error}} shows the formation error $\varepsilon_f$, where $\varepsilon_f=\sum_{i=1}^{n}\sum_{j=1,j\in\mathbb{N}_i}^{n}a_{ij}\left\|\boldsymbol{p}^i-\boldsymbol{p}^j-\delta_{ij}^*\right\|_2$.
\mbox{Fig. \ref{fig:velocity_tracking_error}} shows the norm of the velocity tracking error $\varepsilon_v^i$, defined as $\varepsilon_v^i=\left\|\boldsymbol{v}^i-v^*\right\|_2$.
The results demonstrate that MRACon-based NMPC can control the UAV swarm to fly through complex environments while maintaining formation and a set velocity.
When UAVs are avoiding obstacles or other UAVs, formation and velocity tracking errors are impacted.
 The formation and velocity tracking errors converge once the collision avoidance maneuvers are completed.

\subsection{Hardware Experiment}
\begin{figure}[!htbp]
	\centering
	\includegraphics[width=4.0cm]{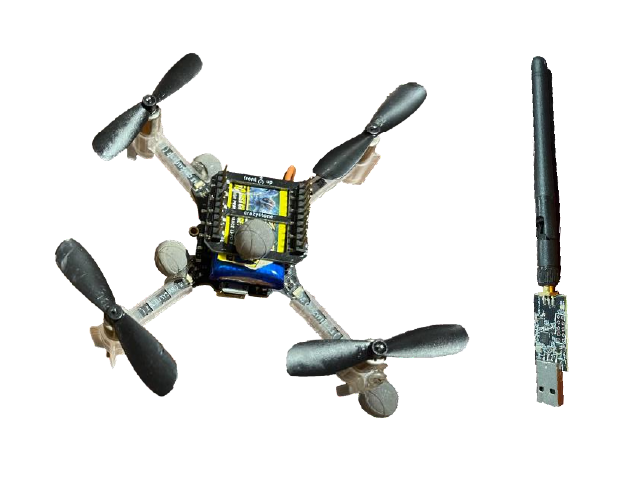}
	\caption{One of the equipment used in the experiment. The left side is the Crazyflie UAV, and the right side is the Crazyradio transceiver.}
    \label{fig:experiment_crazyflie}
\end{figure}

\begin{figure}[!htbp]
	\centering
	\includegraphics[width=9.0cm]{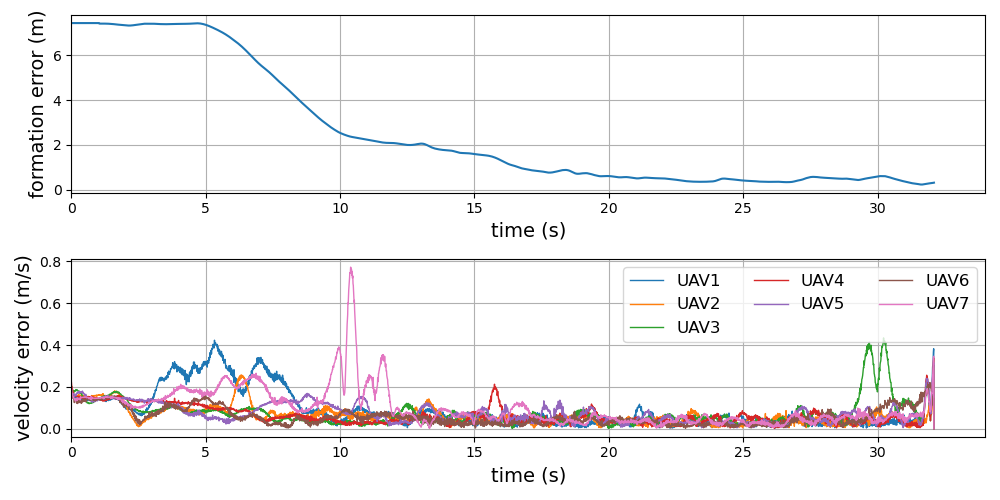}
	\caption{The upper graph shows the formation error $\varepsilon_f$, and the below shows the velocity tracking error $\varepsilon_v^i$ of all UAVs.}
    \label{fig:formation_error_real}
\end{figure}

\begin{figure}[!htbp]
	\centering
	\includegraphics[width=9.0cm]{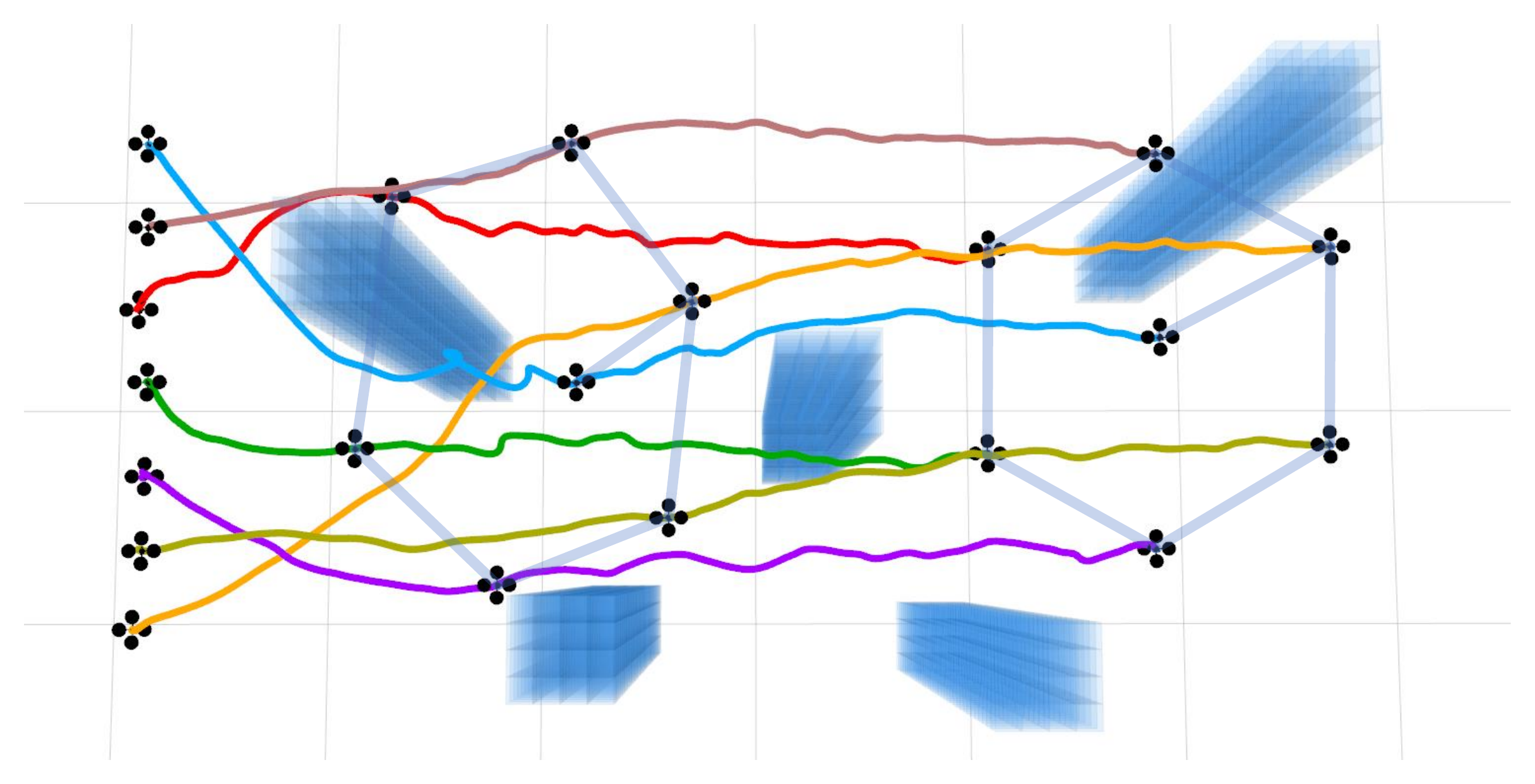}
	\caption{Formation trajectory in the hardware experiment.}
    \label{fig:formation_traj_real}
\end{figure}

\begin{figure}[!htbp]
	\centering
	\includegraphics[width=9.0cm]{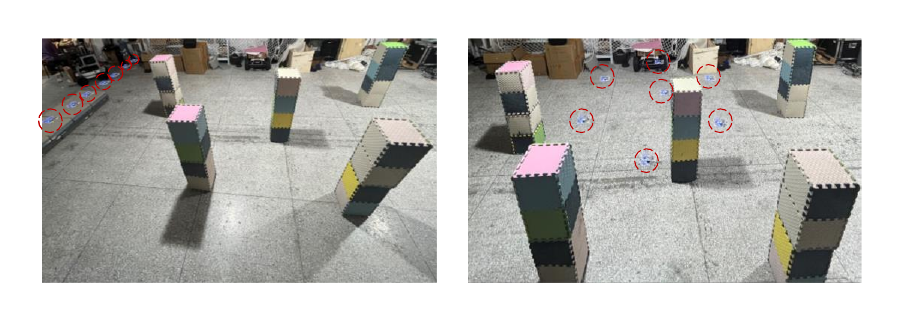}
	\caption{Formation scenario in the hardware experiment. From left to right are the formation scenarios at 0s and 28s, respectively.}
    \label{fig:formation_scenario}
\end{figure}

We conduct hardware experiments to further verify the effectiveness of the MRACon-based NMPC.
Our hardware experiment platform consists of seven Crazyflie UAVs, a Vicon motion capture system, and a laptop computer with three Crazyradio transceivers. 
To establish communication between the laptop and UAVs, we use a ROS package developed for Crazyflie \cite{honig2017flying}.
States of the Crazyflie and the obstacle information are acquired by the Vicon system and the onboard IMU.
\mbox{Fig. \ref{fig:experiment_crazyflie}} shows the UAV and the transceiver used in the experiment.
\mbox{Fig. \ref{fig:formation_error_real}} demonstrates the formation error and velocity tracking error of all UAVs.
\mbox{Fig. \ref{fig:formation_traj_real}} displays the trajectories of the UAVs in the formation.
\mbox{Fig. \ref{fig:formation_scenario}} shows three formation scenarios in the hardware experiment.
It can be seen that a hexagon-shaped formation has been achieved, and the UAVs are maneuvering at a set velocity.
\section{Conclusion}
In this paper, we have presented a distributed MRACon-based NMPC algorithm that enables UAVs to achieve formation flying with collision avoidance. 
The formation algorithm of the second-order reference model, tracking cost function, obstacle avoidance, and reciprocal avoidance constraints 
were designed in the algorithm. Moreover, we have analyzed the stability of the proposed algorithm.
Finally, the algorithm's effectiveness has been verified by simulation experiments and hardware experiments.
In future work, we will focus on studying the problem of leaderless UAV swarms flying automatically in the forest environment with an onboard perception system.

\bibliographystyle{ieeetr}
\bibliography{export}

\end{document}